# Bell state generation and CNOT operation using on-demand identical photons from shape-controlled spatially ordered quantum dots

**Qi Huang, Swarnabha Chattaraj[1], Lucas Jordao, Jiefei Zhang[1], Siyuan Lu[2] & Anupam Madhukar**

*Nanostructure Materials and Devices Laboratory, University of Southern California,*

*Los Angeles, CA 90089-0241*

Nov. 6, 2024

**Abstract**:

Fault tolerant on-chip photonic quantum computation is enormously helped by (a) deterministic generation of the needed thousands to millions of photon qubits from (b) quantum emitters in designed spatially ordered arrays to enable networks for implementing many-qubit logic circuits. Scaling up photonic quantum information processing systems has, however, been prevented by the lack of such quantum emitters until the demonstration of the platform of mesa-top single quantum dots (MTSQDs) – controlled shape, size, and volume single QD- located in designed regular arrays. Here we demonstrate 2 qubit CNOT gate operation- a universal gate necessary to enable quantum circuits of arbitrary complexity- in polarization basis using photons emitted from individual MTSQDs. A Bell state fidelity of $0.825\pm0.010$ is achieved with two photon interference (TPI) visibility of $0.947\pm0.0015$ at 4K *without* Purcell enhancement. The results make a strong case for developing MTSQD arrays for utility scale optical quantum information processing platforms.

## Introduction

Quantum information processing (QIP) has the potential to revolutionize many fields, including rapid speed up in solving certain computational tasks [1] and secure communication [2]. Many different types of qubits are being explored for QIP including atoms [3], ions [4], photons [5, 6], defect deep levels [7], and superconducting Josephson junctions-based circuits [8]. Among these, photon qubits used in optical quantum information processing have certain advantages such as weak interaction with the environment, long coherence time, room temperature operation, and fast travel speed that enables rapid connection between remote nodes. For universal QIP, a universal set of quantum gates is needed that includes the CNOT gate which can entangle two separate qubits. However, while the weak interaction between photons and the environment is beneficial for maintaining coherence, it poses a significant challenge for the CNOT gate implementation due to the absence of natural photon-photon interactions. The required nonlinearity, shown by Knill, Laflamme and Milburn [9], can be achieved using only linear operations by introducing detection- resulting in a universal scheme for linear optical QIP. Though the CNOT gate operation can be achieved with linear optics, the success probability is limited to 1/9 using two un-entangled single photons [10]. Other multi-photon qubit operations also face challenges due to the probabilistic nature of the gate operations. Overcoming these limitations requires an increased number of single-photon sources (SPSs) and more complex circuit architectures [9]. Furthermore, to enable fault tolerance, estimates of the number of physical qubits required to implement a logical qubit range from $10^3$ to $10^6$ [11]. Consequently, scaling up to practical applications involving thousands of logical qubits would demand millions of physical qubits. The needed size of systems for utility-scale QIP is clearly out of reach of the current pathway of using time-delayed photon from a single or few SPSs [12] that can at best scale up to ~100 photons emitted in sequence [13]. Thus, the need for precise spatial placement of arrays of SPSs at pre-designed locations compatible with circuit layouts is indispensable for achieving practical photonic QIP. In addition, QIP systems demand interconnection between various SPSs and functional nodes. All these point, yet again, to the need for highly accurate positioning of large numbers of SPSs to facilitate the needed scalable integration.

---

1 Currently at Argonne National Laboratory. Illinois. 60439, United States.
2 Currently at XPENG. California. 95054, United States.

The requirement of a spatially ordered array of SPSs in on-chip photonic QIP is independent of the particular application purpose such as Boson sampling circuits for quantum simulation [14], or the KLM protocol-based [9] and measurement-based quantum computation (Fusion) [15] paths to computation. The general structure of the on-chip photonic system can be captured symbolically as in Fig. 1(a); it involves an SPS array feeding into light manipulating circuitry that performs unitary operations to create and process the desired N-photon entangled state using building blocks (units) of Hong Ou Mandel (HOM) and Mach Zehnder interferometers (MZIs) with output photons guided into a co-designed array of detectors (preferably photon number resolving detectors).

We have established [16-20] that arrays of mesa-top single quantum dots (MTSQDs) provide the only platform with sufficiently accurately spatially ordered location and spectrally uniform, on-demand, SPSs needed for creating networks aimed at quantum information processing. The MTSQDs generate highly pure single photons ($g^2(0) < 0.01$) [17, 18] with, as shown here, a fast decay life $T_1$~350ps - 400ps *without* any Purcell enhancement. This short decay time of as-synthesized MTSQDs corresponds to an oscillator strength of $f$~27 - 29 which is over 3 times larger than other types of typical SQDs (single quantum dots) [21,22]. This large oscillator strength, we show here, is attributable to the larger volume of MTSQDs enabled by the control provided by the SESRE (substrate-encoded size-reducing epitaxy) growth method [16,17,23,24]. As captured in Fig.1(a), using the MTSQD platform, large numbers of spectrally uniform SPSs can be integrated, monolithically, or hybridized with SOI-based photonics providing the on-chip photon manipulating circuitry (marked Linear Optical Circuits in Fig.1(a)) to create the needed large-scale structures without resorting to the impractical individual pick-and-place approach [25]. The needed array of SPSs can be provided by a single transfer of an active membrane containing the array of MTSQDs in predesigned locations- a capability missing in other classes of QDs and deep level SPSs.

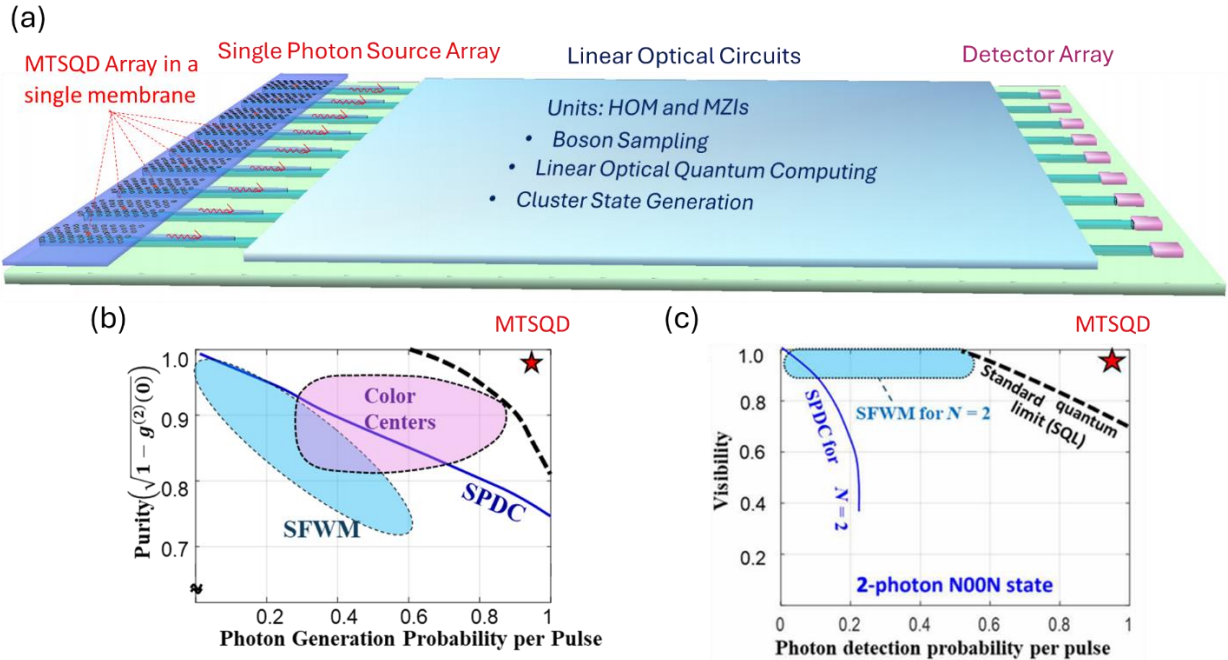

**Figure 1 (a).** Schematic illustrating large scale on-chip photonic QIP system enabled by the MTSQDs providing an array of on-demand SPSs that can be integrated with on-chip linear optical circuit with a single integration step in arbitrary scale. **(b)** and **(c)** (adopted from [17]) indicate, respectively, the theoretical analyses-based required values (above the black broken lines) of single photon purity and two-photon interference visibility for suitability of any quantum emitter for efficient quantum information processing. The red star indicates MTSQD.

To guide the path to creating large optical circuits as envisioned in Fig.1(a), in this work we first establish the 4K two-photon interference visibility as a function of temperature and the time delay between subsequent photon emitted from the MTSQD, a figure-of-merit that underpins the higher level (two and more qubits) functions, such as the CNOT gate, needed for creating circuits. At 4 K, a two-photon interference (TPI) visibility of ~0.947 ± 0.005 is achieved for a 2 ns delay which remains a high ~0.80 at 20 K. Using the high-purity and indistinguishable photons emitted from a single MTSQD, we demonstrate a CNOT gate operation with a 1/9 success probability. A maximally entangled Bell state is generated with the CNOT gate, achieving a state fidelity of $0.824 \pm 0.010$ extracted from a full quantum state tomography. The success probability is limited by the inherent probabilistic scheme used here in linear optical system. It could be further improved with added resources [26], including number resolving photon detectors and ancilla photons with dual-rail or time bin encoding. Our results suggest that MTSQDs fabricated using the SESRE method is a promising platform of scalable SPSs towards photonic quantum information processing where large number of physical qubits are needed.

## Results

### MTSQD Large Volume Control & Resulting Large Oscilator Strength

The planarized 4.25 ML $In_{0.5}Ga_{0.5}As$ MTSQDs (Fig. 2(c)) are arranged in multiple (M × N) arrays. The MTSQDs are positioned on top of a DBR mirror consisting of 17.5 pairs of $\lambda/4$ thick AlAs and GaAs. The DBR mirror is designed to enhance photon collection efficiency at the first objective lens by a factor of approximately 10, without introducing Purcell Enhancement [17]. Details of the growth can be found in Ref.[17] and the methods section. MTSQDs position accuracy (~5 nm) is determined by the accuracy of electron beam lithography patterning and wet chemical etching of the starting pedestal mesa (Fig. 2(a) and (b)). For the MTSQDs in 5×8 array, the neutral exciton emission wavelength standard deviation is found to be ~2.8 nm as reported in [17].

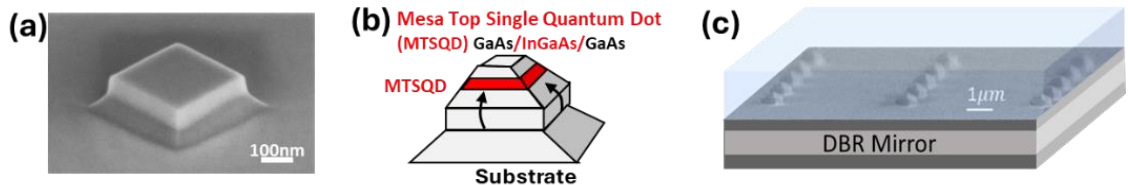

**Figure 2.** **(a)** SEM image of a typical pedestal nanomesa fabricated in spatilly ordered arrays using electron beam lithography and wet chemical etching; which is used for MTSQDs growth using SESRE approach, **(b)** schematic of SESRE growth showing cation adatom migration from side wall to mesa top and enabling selective formation of a SQD (red region) on the mesa top; **(c)** composite of a SEM image (grey) showing an array of as-formed pyramidal structures that contain MTSQDs before surface morphology planarization and schematic of the GaAs planarization layer (translucent blue layer). The MTSQDs are sitting on top of a DBR mirror incorporated in the substrate to enhance the photon collection efficiency.

The integrated and time-resolved emission characteristics of neutal exciton emission from MTSQDs were investigated as a function of excitation laser power and sample temperature. The sample was mounted in a cryogen-free cryostat and excited resonantly using normal incidence through a 40X NA 0.6 objective lens. We used a confocal microscope with cross-polarization configuration to enable back (normal) detection of the emisison. The excitation laser was operated at a repetition rate of 76 MHz, with a pulse width of approximately 3 ps, and the scattered excitation laser light is suppressed with an extinction ratio of ~$10^7$ by the cross-polarization confugraion microscope. Further details of the optical measurement setup are provided in the Methods section.

Figure 3 (a) shows the detected resonant photoluminescence (PL) count rate of the neutral exciton emission from a typical MTSQD as a function of the square root of the excitaiton laser power. The gray dashed line serves as a visual guide. The meausred data exhibit the well known Rabi oscillation, indicating coherent manipulation between the ground state $|0\rangle$ and one neutral exciton state $|1\rangle$ in the MTSQD.

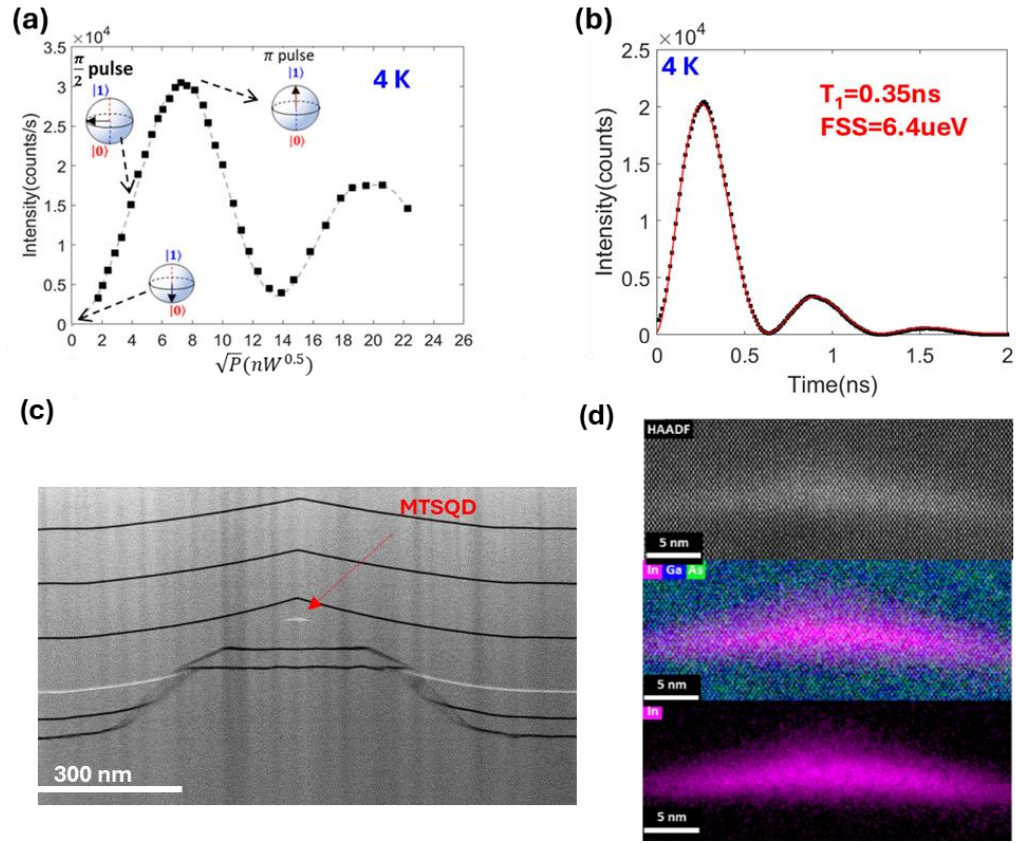

**Figure 3. (a)** Measured resonant photoluminescence intensity as a function of square root of laser power, showing clear Rabi oscillation. **(b)** measured time-resolved photoluminescence (TRPL) from one MTSQD with resonant excitation at $\sim\pi/2$ pulse (power 19.6 nW, 1.6 W/cm$^2$) at 4 K. The red curve is a fitting of the measured data. **(c)** Cross-sectional Z-contrast TEM image revealing the growth front evolution during SESRE growth. The small white region centered on the mesa-top corresponds to InGaAs region surrounded by GaAs (grey region), the two together defining the MTSQD. The dark lines are the thin AlGaAs marker layers interspersed to mark the growth profile; **(d)** STEM image of an *as grown* planarized MTSQD. The uppermost panel shows the high-angle annular dark-field (HAADF) scanning spectroscopy image of the MTSQD. Middle panel shows the energy dispersive X-ray spectroscopy (EDS) image of the same MTSQD highlighting Indium, Gallium and Arsenic. The lowermost panel shows the EDS image of Indium distribution. Note the well-defined shape and large volume (base ~30 nm, height ~5.5 nm).

The neutral exciton decay lifetime is obtained from the time resolved photoluminescence (TRPL) measurements where a 15 $\mu$eV resolution spectrometer is employed to further suppress the scattered laser light. Figure 3 (b) shows the measured TRPL at an excitation power of 19.6 nW, 1.6 W/cm$^2$ (near $\sim\pi/2$ pulse). The observed oscillatory behavior within the exponential decay envelope is attributed to the self-interference of the photon wave packet, which is a coherent superposition of two fine structure split states. The behavior can be described by [17]:

$$I(t) \propto \left| \exp\left(-i\Delta t - \frac{t}{2T_1}\right) - \exp\left(-\frac{t}{2T_1}\right) \right|^2 \tag{1}$$

where $\Delta$ denotes the fine structure splitting energy and $T_1$ the decay lifetime. The red curve shows the fit to the data, revealing a decay lifetime $T_1$ of ~350 ps and fine structure splitting ~6.4 $\mu$eV. It is worth noting that for typical InGaAs/GaAs SAQDs (self-assembled quantum dots), the decay lifetime is ~1 ns in the absence of Purcell enhancement. The MTSQDs $T_1$ being ~3 times shorter (without Purcell enhancement) indicates a significantly larger oscillator strength for the MTSQD. The oscillator strength $f$ is defined as the ratio between the emission rate of the MTSQD compared to an ideal harmonic oscillator and can be determined as [21]

$$f = \frac{6\pi\epsilon_0 m_0 c^3}{n T_1 F_p \omega^2 e^2} \tag{2}$$

where $m_0$ is the mass of electron, $n$~3.5 is the refractive index of GaAs, $F_p$ is the Purcell factor, $\omega$ is the angular frequency of the emitted photon, and $\epsilon_0$ is the permittivity of free space. Using eqn. (2) the TRPL data of Figure 3(b) gives f ~29, a value 3-4 times the oscillator strength for typical SAQDs of ~7-8.

The observed large oscillator strength, we suggest, is a result of a superradiant effect known to kick in for large quantum dot (here InGaAs) volumes [27] for which the exciton is nearly wholly contained inside, and its center of gravity weakly bound. The exciton thus samples a significantly larger number of Bloch unit cells which defines the operational oscillator strength. Indeed, our STEM studies of these MTSQDs, summarized in Fig.3 (c) and (d), provide confirming evidence. The Z-contrast STEM image of Fig.3 (c) reveals the growth front profile evolution via the AlGaAs marker layers (dark lines) in otherwise GaAs (grey region) growth and shows the placement of the InGaAs quantum dot region (the white region) centered on the size-reduced mesa top. Figure 3(d), uppermost panel, is an atomic-resolution high-angle annular dark field (HAADF) scanning transmission electron microscope (STEM) based image of the MTSQD region revealing defect-free InGaAs SQD. The middle and lowermost panels of Figure 3(d) show the atomic resolution In, Ga, and As composition distributions, as revealed by energy dispersive spectroscopy (EDS) of characteristic X-ray emission from these elements on the same MTSQD, mapped using the In L$\alpha$1, Ga K$\alpha$1 and As K$\alpha$1 X-ray lines. STEM and EDS data were taken on a probe-corrected Spectra 200 X-CFEG STEM instrument, employing an electron beam diameter of 0.8 A, beam current of 100 pA, beam energy of 200 keV, and probe semi-angle of convergence of 25mrad. The HAADF images were acquired with detector collection angle of ~60-200 mrad, while the EDS mapping was done using Dual-X detector with solid angle of ~1.8 sr. The HAADF and EDS results on the MTSQDs confirm the large volume of the grown MTSQDs, consistent with the observed short optical lifetime $T_1$ and the large oscillator strength. Further, we recall that large QD volume typically implies weaker electron-phonon coupling strength, resulting in MTSQDs being robust against phonon induced dephasing process. Indeed, this too is confirmed as discussed below.

**MTSQD Single Photon Purity and Indistinguishability**

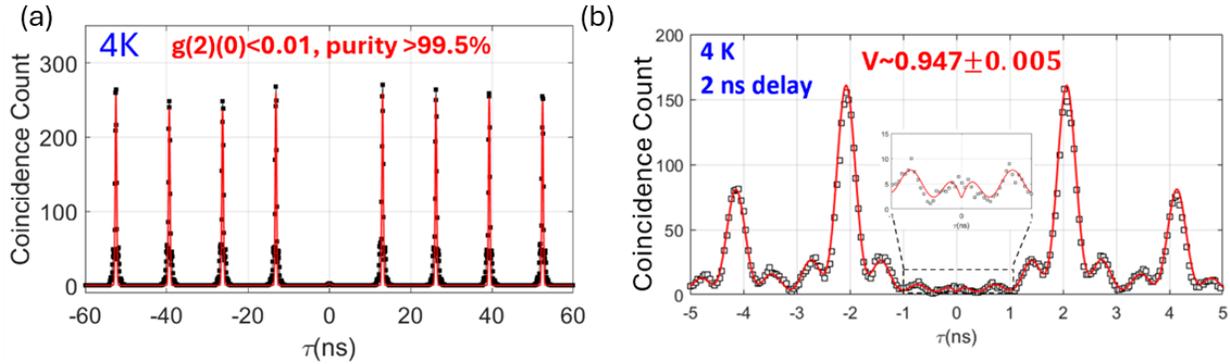

**Figure 4 (a)** Measured coincidence counts histogram from resonantly excited MTSQD in 5×8 array using HBT setup, showing a $g^{(2)}(0)<0.01$ and single photo purity >99.5%. **(b)** Measured coincidence counts histogram of TPI using HOM interferometer.

The measured second order correlation $g^{(2)}(\tau)$ using the HBT (Hanbury-Brown Twiss) setup for the neutral exciton emission under pulsed resonant excitation from the MTSQD is shown in Figure 4 panel (a). By calculating the ratio of the $\tau = 0$ peak area and the average area of $\tau \neq 0$ peaks, the data suggest a $g^{(2)}(0) < 0.01$ and single photon purity > 99.5%, showing highly pure single photon emission from MTSQD.

Single photon indistinguishability is another key figure of merit of quantum emitters for QIP applications. The indistinguishability of the single photons from the MTSQDs was assessed through two-photon interference (TPI) measurements using a HOM (Hong-Ou-Mandel) interferometer (see Methods). Figure 4 (b) shows the measured coincidence counts as a function of time difference ($\tau$) between two consecutive detection events from the two detectors at the output of the HOM interferometer with the excitation laser pulse separation $\delta t$=2ns. The data shown in Figure 4 (b) is corrected for laser leakage under resonant excitation. The near absence of coincidence counts at $\tau \sim 0 \ ns$ indicates a high TPI interference visibility, revealing a TPI visibility $V_{2ns} = 0.947 \pm 0.005$ for the MTSQD. The peaks that are ~±0.6 ns away from the main peaks at $\tau = 0 \ ns, \pm 2 \ ns,$ and $\pm 4 \ ns$ in the coincidence counts data are also a result of photon self-interference as discussed in the TRPL data. Note that the expected dip in the photon coincidence count at $\tau = 0 \ ns$ [17] is not prominent in the measured data since its width (~80 ps) is comparable to our instrument response. This high visibility is also observed in MTSQDs grown in a 50×50 array, showing that the MTSQDs can be scaled to large area arrays while maintaining similar characteristics [20].

As indicated by the TRPL results, the MTSQDs show a relatively large oscillator strength ($f \sim 29$), which is likely due to their substantial volume. Typically, QDs with a large volume couple to phonons weakly [28]. Thus, MTSQDs with large volume are expected to be less affected by phonon-induced dephasing. This hypothesis is tested by measuring the two-photon interference (TPI) visibility as a function of the sample temperature. Figure 5 (a) shows the TPI visibility as a function of substrate temperature from 4 K to 40 K from the MTSQD with a 2 ns pulse separation. The TPI visibility can hold to ~0.80 at 20 K and a moderate value of ~0.45 at 40 K, showing a slower reduction of visibility as temperature is increased compared with typical SAQDs [29-32].

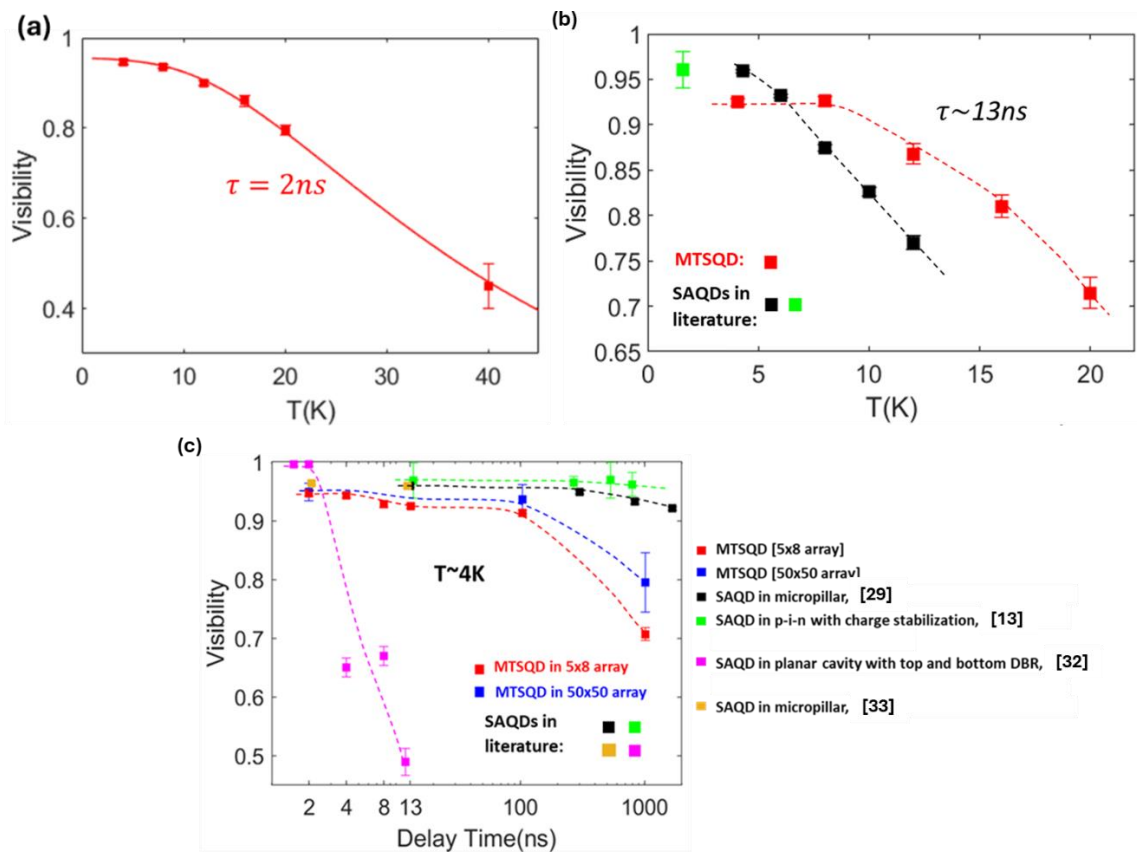

**Figure 5. (a)** Measured TPI visibility as a function of temperature for a constant 2 ns delay between the photons emitted from the MTSQD in 5×8 array with $\pi/2$ pulse excitation. The solid line is a fitting to the measured results following eqn. (3). **(b)** Two photon interference visibility as a function of temperature for the more standard 13 ns delay and comparison with SAQDs. MTSQDs, being large in size, are found less sensitive to phonons and hence thermal dephasing processes, and they show visibility > 90% up to 10K and show a much smaller reduction in visibility with increasing temperature compared to SAQDs [data on SAQDs taken from Refs. 13, 29, 32, 33]. **(c)** measured TPI visibility as a function of the time delay $\delta t$ up to ~1 μs.

**Intrinsic Dephasing**

Phonon-induced dephasing in quantum dots primarily arises from two mechanisms: (i) a rapid non-Markovian dephasing process which contributes to the formation of phonon sideband; (ii) a phonon-assisted virtual transition to higher energy states in the QD, which results in pure dephasing of the zero-phonon line (ZPL). Considering these two contributions to phonon-induced dephasing, the TPI visibility can be described by the following expression [34]:

$$V = \frac{\Gamma/2}{\Gamma/2 + \gamma_{ph} + \gamma_{sd}} \left[ \frac{B^2}{B^2 + F(1-B^2)} \right]^2 \tag{3}$$

where in the first term, $\Gamma = \frac{1}{T_1}$ is the radiative recombination rate; $\gamma_{ph}$ is the dephasing rate due to virtual phonon transition; and $\gamma_{sd}$ denotes the spectral diffusion induced dephasing. The second term describes the dephasing due to photon emission from the phonon side band, in which $B = \exp\left(\frac{\alpha}{2}\int_0^\infty \nu \exp(-\nu^2/\nu_c^2)\coth(\frac{\nu}{2k_bT})\right)$ is the Franck-Condon factor, and $F$ is the fraction of phonon sideband that is not filtered. The term of virtual phonon transition induced dephasing is $\gamma_{ph} = \frac{\alpha^2\mu}{\nu_c^4}\int_0^\infty \nu^{10}\exp\left(-\frac{\nu^2}{\nu_c^2}\right)n(\nu)[n(\nu)+1]d\nu$, in which $\alpha$ is related to electron-phonon coupling strength; $\mu$ is a parameter associated with the probability of virtual phonon processes (proportional to the inverse of the energy spacing between the ground and excited states of the quantum dot), and $\nu_c$ is the phonon cutoff frequency that is inverse to the QD's confinement length (QD size).

The spectral diffusion is a relatively slow process. With 2 ns pulse separation, the effect of spectral diffusion can be ignored to analyze the temperature dependence result. The red solid line in Figure 5 (a) is the fitted result of TPI visibility versus temperature, with fitting parameters $\alpha = 0.0055 \text{ ps}^2$, $\nu_c = 4.9 \text{ ps}^{-1}$, $\mu = 2.2 \times 10^{-3} \text{ ps}^2$, and $F = 0.3$. To compare with the SAQD literature, we also show in Fig. 5(b) the TPI visibility as a function of temperature at a 13 ns delay. Compared with typical reported [29, 31] SAQDs, the fitted results show that these MTSQDs have a weaker electron-phonon coupling strength and a (controlled) larger volume, consistent with the observed large oscillator strength (Fig. 3(b)) that pointed to a large MTSQD volume as shown in Fig. 3(d).

The larger degree of isolation from the phonon-induced dephasing in MTSQDs compared to SAQDs implies that in the MTSQD array based photonic QIP, the number of resource single photon states can be vastly increased by using a chain of single photons emitted from each MTSQD in the array. Since the number of single photons that can be used from each MTSQD is limited by the spectral diffusion timescale, we here investigate this timescale for our MTSQD platform. To determine the time scale of spectral diffusion, TPI visibility was measured as a function of excitation pulse separation $\delta t$. As shown in Figure 5 (c), the visibility remains V~0.91 with pulse separation up to $\delta t$~105 ns, suggesting that with a comfortable ~200 MHz excitation frequency (5 ns separation), a single MTSQD can generate a string of over 20 indistinguishable photons. Note that this result is obtained from a MTSQD sample without employing a p-i(QD)-n configuration to stabilize the charge environment which has been shown to enhance the spectral diffusion timescale for SAQDs (green markers in Fig. 5(c)). We also note that by including cavity structures around each MTSQD, as represented in Fig.1, a further enhancement of the TPI visibility is expected. A representative result for SAQD in micropillar cavity shown as the black markers in Fig. 5(c) illustrates this point. When the pulse separation is extended to $\delta t \approx 1\ \mu s$, the visibility is currently limited to approximately 0.71. From [30], dephasing from spectral diffusion can be described by $\gamma_{sd} = \Gamma_{sd}\left(1 - e^{-(\delta t/\tau_c)^2}\right)$, where $\tau_c$ is the correlation time scale of the spectral diffusion, extracted from the fitting as approximately 350 ns. This correlation time is significantly longer than that observed in typical SAQDs without charge stabilization [30, 32]. Given that charge fluctuation induced dephasing is related to the density of random fluctuating charge traps near the MTSQDs, the spectral diffusion can be minimized by improving the material quality and stabilizing the charge environment with p-i-n structure.

**CNOT gate behavior and Bell state generation.**

The above findings suggest that in the MTSQD array based photonic QIP circuits we can in principle harvest multiple single photons emitted from each MTSQDs in the array to potentially create the needed millions of photon-based resource-qubits for utility scale applications. As a first step towards this direction, we here demonstrate a controlled-NOT (CNOT) gate operation using photons from the same MTSQD. The CNOT gate is a fundamental two qubit logic gate which is required for universal quantum information processing. For CNOT gate operation, if the control qubit is $|0\rangle_c$, the target qubit remains unchanged. However, if the control qubit is $|1\rangle_c$, the target qubit will then be flipped: $|0\rangle_t \rightarrow |1\rangle_t$, $|1\rangle_t \rightarrow |0\rangle_t$. The experimental setup for the CNOT gate is shown in Figure 6 (a), in which the quantum information is encoded in the polarization degree of freedom: $|H\rangle = |0\rangle$ and $|V\rangle = |1\rangle$. The key element in the CNOT gate, implemented using linear optics, is a set of three partially polarizing beamsplitters (PPBS) that together act as a controlled-phase (CZ) gate in the coincidence detection basis [37]. The central PPBS transmits horizontally polarized photons, with transmission efficiency $T_H$ =1, but only partially reflects (transmits) vertically polarized photons, with reflection efficiency $R_V$ =2/3 ( $T_V$ =1/3). When two indistinguishable photons in the $|VV\rangle$ state simultaneously arrive at the central PPBS, the Hong-Ou-Mandel effect introduces a $\pi$ phase shift if the photons exit from separate ports (which contribute to the coincidence detection). For other input states ($|HH\rangle$, $|HV\rangle$, $|VH\rangle$), there is not two photon interference at the central PPBS as H-polarized photon has perfect transmission. The two PPBSs after the central PPBS are each rotated by 90° with respect to the central PPBS, giving H polarized photons a transmission of $T_H$=1/3, so that the amplitude of H polarized component matches with the $|VV\rangle$ state. CZ gate implemented in this configuration has an overall success probability of 1/9. When combined with two half-wave plates (HWPs) set at 22.5° to act as Hadamard gates on the target qubit, one forms the CNOT gate, as highlighted in the red region of Figure 6(a).

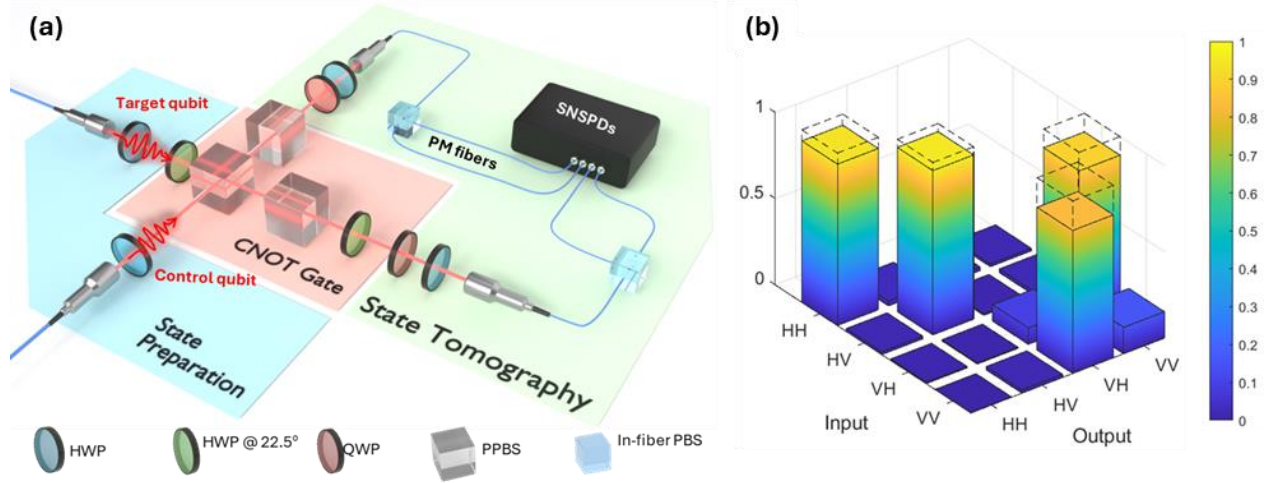

**Figure 6.** **(a)** the CNOT gate setup consisting of three regions (1) state preparation to set the photon to needed polarization, (2) the CNOT gate, and (3) state tomography to project photons to different polarization basis. **(b)** the measured truth table of CNOT gate operation in ZZ basis (dashed lines indicate ideal values).

The CNOT gate performance was evaluated by measuring the probabilities of detecting different output states for four mutually orthogonal input states. Figure 6 (b) shows the CNOT gate operation truth table in the ZZ basis (where $|H\rangle/|V\rangle$ are the eigen states). For ideal CNOT gate operation in ZZ basis, the target qubit flips polarization when control qubit is in $|V\rangle$, and the ideal CNOT gate operation is marked as black dashed line in Figure 6 (b). The CNOT gate operation fidelity in the ZZ basis is defined as $\mathcal{F}_{ZZ} = \frac{1}{4}[P(HH|HH) + P(HV|HV) + P(VH|VV) + P(VV|VH)]$, where $P(I_1I_2|O_1O_2)$ represents the probabilities of detecting output state $|O_1O_2\rangle$ given the input state $|I_1I_2\rangle$ from the CNOT operation. From the truth table extracted from these projection measurements of Figure 6 (b), the estimated fidelity is $\mathcal{F}_{ZZ} = 0.902 \pm 0.043$.

The overall quantum process fidelity ($\mathcal{F}_{proc}$) [38] of the CNOT gate can be shown to be bounded as $(\mathcal{F}_{XX} + \mathcal{F}_{ZZ} - 1) \leq \mathcal{F}_{proc} \leq \min(\mathcal{F}_{XX}, \mathcal{F}_{ZZ})$ where $\mathcal{F}_{XX}$ and $\mathcal{F}_{ZZ}$ represent fidelity in two complementary basis (here X and Z). Thus, to estimate $\mathcal{F}_{proc}$, we also evaluate the CNOT gate fidelity in XX basis (eigen states $|D\rangle/|A\rangle$) in a similar manner of that in ZZ basis. The fidelity in the XX basis is found to be $\mathcal{F}_{XX} = 0.874 \pm 0.041$, setting the bounds of the quantum process fidelity as $0.776 \leq \mathcal{F}_{proc} \leq 0.874$, which exceeds 0.5. This confirms the entangling capability of the CNOT gate.

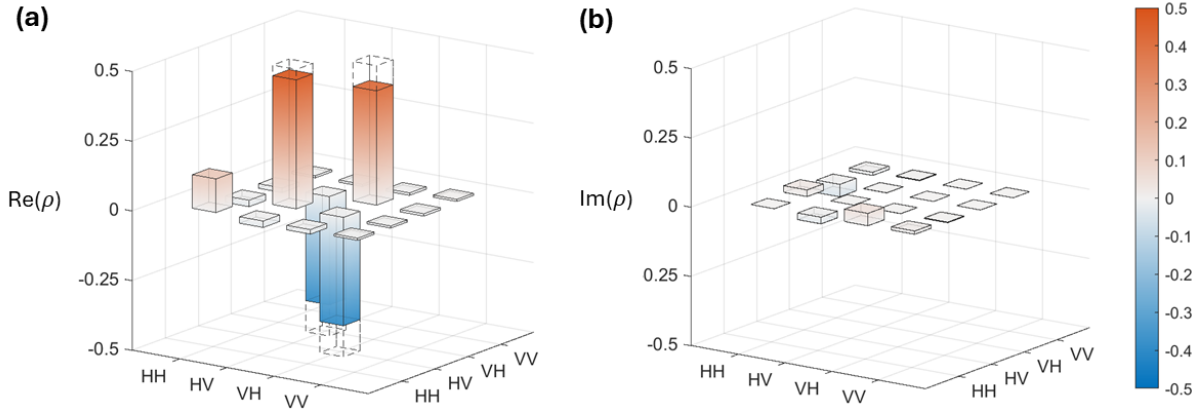

**Figure 7.** **(a)** real and **(b)** imaginary parts of the density matrix of the Bell state generated by the CNOT gate. The density matrix of the ideal Bell state $|\Psi^-\rangle$ is marked with black dashed bars in (a).

We proceed to generate a maximally entangled Bell state using the CNOT gate starting from a pair of independent photons from the same MTSQD. By setting the control qubit as $|A\rangle = 1/\sqrt{2}(|H\rangle - |V\rangle)$ and target qubit set as $|V\rangle$, a Bell state $|\Psi^-\rangle = \frac{1}{\sqrt{2}}(|HV\rangle - |VH\rangle)$ is generated at the output of the CNOT gate. To fully characterize the generated state, we performed a full quantum state tomography [39], involving 4×9=36 coincidence measurements across the combinations of $|H\rangle, |V\rangle, |D\rangle, |A\rangle, |R\rangle$ and $|L\rangle$ staes using four superconducting nanowire detectors. A maximum likelihood estimation algorithm is then performed based on the 36 coincidence measurements to reconstruct the density matrix $\rho$. Figure 7 (a) and (b) shows the real and imaginary part of the density matrix. From the measured density matrix, we estimate the degree of entanglement, quantified by concurrence and Von Neumann entropy (entanglement entropy), as $C = 0.745 \pm 0.021$ and $S = 0.797 \pm 0.036$ respectively, indicating entangled nature of the generated state. We test its closeness with the ideal Bell state using the metric of state fidelity defined as $\mathcal{F}_{\Psi^-}(\hat{\rho}_{ideal}, \hat{\rho}) = \left(\text{Tr}\{\sqrt{\sqrt{\hat{\rho}_{ideal}}\, \hat{\rho}\, \sqrt{\hat{\rho}_{ideal}}}\}\right)^2$. We estimate $\mathcal{F}_{\Psi^-} = 0.825 \pm 0.010$ from the reconstructed density matrix. The error bar here is determined using Monte Carlo simulations, assuming Poisson-distributed measurement uncertainties, and reconstructing the density matrix to calculate the corresponding fidelity. The state fidelity is well above the required value of 0.78 for CHSH violation assuming an isotropic noise channel [40], indicating that the generated Bell state can serve as a quantum resource for QIP.

**Outlook / Discussion.**

A 2-qubit CNOT gate represents the smallest nontrivial "quantum circuit" and has been used as a steppingstone to development of larger circuits in many platforms. In this work we demonstrate this fundamental CNOT operation for the MTSQD SPSs and Bell state generation at 4K *without* any Purcell enhancement or charge stabilization techniques. Compared to SAQDs, the MTSQDs grown in spatially-selective locations using the SESRE approach are shown to have controlled shape, size, and enabling large volume which makes them robust against acoustic phonon-induced dephasing. This results in a smaller overall dephasing rate and large two-photon visibility as a function of the time delay between subsequently emitted photons. We show that without any charge stabilization techniques the 4K TPI visibility V~0.95 at 2ns time delay $\delta t$ remains V~0.91 up to $\delta t$~105 ns. Further increase of this correlation time is expected by employing p-i-n structure, to suppress charge noise, and by employing photonic cavity structures. As a demonstration of the functionality of such individual MTSQDs, we report an implementation of CNOT gate in the polarization basis using time-delayed photons from the same MTSQD. The results indicate a CNOT gate operation fidelity in ZZ basis > 0.9 and an overall quantum process fidelity > 0.776, testifying the quantum nature. Using the CNOT gate we further demonstrate creation of maximally entangled Bell states with fidelity of 0.825. One could expect to further improve the CNOT operation fidelity and Bell state fidelity with added cavity and charge stabilization control in p-i-n type structure.

MTSQDs placed in designed spatial locations to an accuracy of ~3nm laterally and <1nm vertically [17, 19] across large areas combining time-delayed photon trains from desired numbers of MTSQDs thereby providing a scalable route to the millions of single photon resource states needed for utility-scale photonic QIP as depicted in Fig. 1 (a). As examples of specific application, in Fig. 8 we show two schematic circuits that represent (a) an arbitrary unitary operation on N-qubit state enabling large-scale tunable Boson sampling, and (b) on-chip probabilistic GHZ state generation that provides the building block towards formation of large 3D cluster states for fusion-based quantum computing. In both cases, MTSQD arrays can allow the necessary scale up currently absent in other approaches. Large-scale integration can allow hierarchical building up of complexity that is inherent in integrated utility-scale photonic QIP.

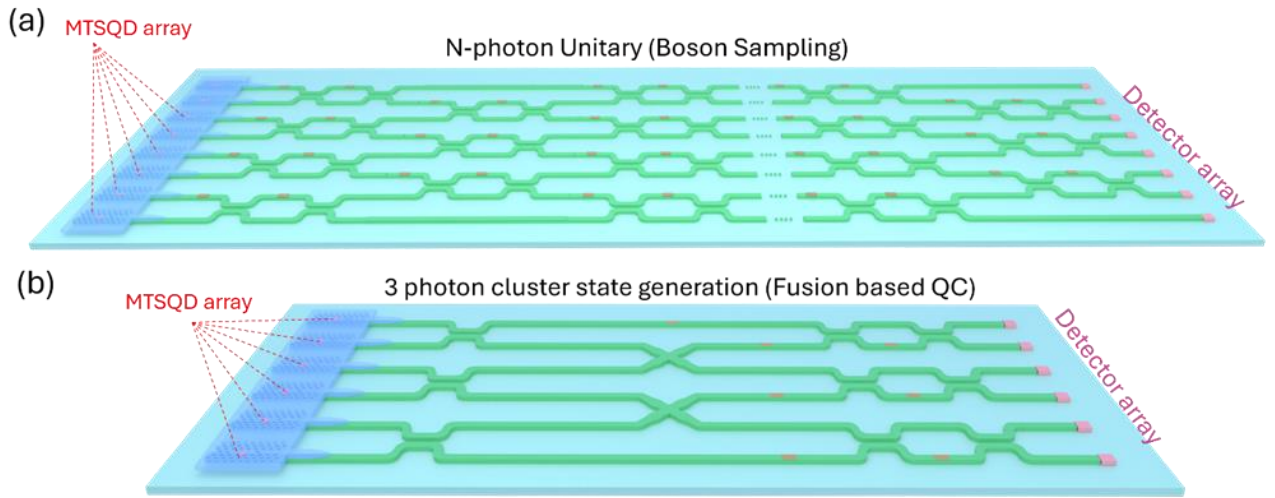

**Figure 8.** Illustrative examples of large-scale on-chip QIP systems that are possible to realize with the MTSQD array SPSs. (a) tunable Boson sampling represented by N-photon unitary operation implemented by a series of Mach-Zehnder Interferometers integrated on-chip with the MTSQD array SPSs, and (b) 3 photon cluster state (GHZ state) generation circuit.

**Acknowledgements**

This work is supported by the Air Force Office of Scientific Research grant number FA9550-22-1-0376, the Center for Nanoimaging (CNI) at the University of Southern California, and the Kenneth T. Norris Professorship.

**Competing interests**

The authors declare no competing interests.

**Methods**

**MTSQD synthesis**

MTSQDs are grown on DBR mirrors using SESRE approach. A DBR mirror consisting of 17.5 pairs of $\lambda/4$ thick AlAs and GaAs is grown first. We then create square pedestal nanomesa oriented along the $\langle 100 \rangle$ crystal direction [18] on the grown DBR mirror using electron beam lithography followed by wet chemical etching for the subsequent growth of MTSQDs. Each pedestal nanomesa has a total height of ~96nm and lateral dimension of ~300nm, located on the GaAs layer above the DBR mirror. The mesa top opening size decreases during the size-reducing growth process due to curvature-induced adatom migration towards the mesa top. After depositing 271ML of GaAs at a growth rate of 4 sec/ML and an arsenic partial pressure $P_{As}$=1.5 × 10$^{-6}$ torr at 600°C, 4.25ML In$_{0.5}$Ga$_{0.5}$As is deposited at 520°C to form MTSQDs at the apex of the nanomesa. The formed MTSQDs are then capped and planarized with an additional total 1346ML of alternating GaAs and AlAs layers.

**Optical Configuration for Resonant measurements**

All single photon emission characteristics data reported in the manuscript has been measured using resonant excitation scheme. The MTSQDs are excited resonantly at their neutral exciton emission wavelengths using a Ti:Sa Mode lock laser with pulses of 3ps width. The scattered laser light is filtered out using, for all measurements (PL, time-resolved PL, HBT, HOM), a cross-polarization configuration (using two polarizers and one polarizing beam splitter). The exciting laser electric field is along the [110] direction with an accuracy of ±3° and the photons with polarization along [-1 1 0] direction are detected, also with accuracy of ±3°. A cross-polarized extinction ratio >1×10$^7$ is established for the resonant excitation studies reported here. The collected photons from the MTSQDs are spectrally resolved with a spectrometer with 15μeV resolution to reveal the true emission linewidth from the MTSQD and also to act as high-resolution spectral filter to filter out unwanted scattered laser light. The collected photons are then detected by superconducting nanowire detectors. For HBT measurements, the spectrally filtered photons emitted from the MTSQDs enter the 50/50 beamsplitter of the Hanbury-Brown and Twiss setup and are detected at the transmitted and reflected ports of the beamsplitter using two superconducting nanowire detectors. The instrument response function for the HBT measurement is ~50-100ps. For HOM measurements, we use the Ti-Sa laser to generate pairs of excitation pulses of width ~3ps with a time separation Δt, controlled by an unbalanced Michelson interferometer built on the laser side. The emitted photon from MTSQDs enters the home-built Hong-Ou-Mandel interferometer are detected by the superconducting nanowire detectors. The instrument response function for the HOM measurement is ~50-100ps. More details can be found in Ref. 19.

### Reference

1. P. W. Shor, Polynomial-Time Algorithms for Prime Factorization and Discrete Logarithms on a Quantum Computer. *SIAM Review* **41:2**, 303-332 (1999)

2. R. Gisin, R. Thew, Quantum communication. *Nat. Photonics* **1**, 165–171 (2007).

3. M. Saffman, Quantum computing with atomic qubits and Rydberg interactions: Progress and challenges. *J. Phys. B At. Mol. Opt. Phys*. **49**, 202001 (2016).

4. C. D. Bruzewicz, J. Chiaverini, R. McConnell, and J. M. Sage, Trapped-ion quantum computing: Progress and challenges. *Appl. Phys. Rev*. **6**, 021314 (2019).

5. Quantum Dots for Quantum Information Technologies, Ed. P. Michler (Springer, 2017).

6. F. Flamini, N. Spagnolo, and F. Sciarrino, Photonic quantum information processing: a review, *Rep. Prog. Phys*. **82**, 016001 (2019).

7. S. Castelletto and A. Boretti, Silicon carbide color centers for quantum applications, *J. Phys. Photonics* **2**, 022001 (2020).

8. M. Kjaergaard, M. E. Schwartz, J. Braumüller, P. Krantz, J. I. J. Wang, S. Gustavsson, and W. D. Lover, Superconducting Qubits: Current State of Play. *Annu. Rev. Condens. Matter Phys*. **11**, 369–395 (2020).

9. E. Knill, R. Laflamme and G.J. Milburn. "A scheme for efficient quantum computation with linear optics", *Nature* **409**, 46 (2001).

10. W. Q. Liu and H. R. Wei, Linear Optical Universal Quantum Gates with Higher Success Probabilities, *Adv Quantum Technol* **6**, 1 (2023).

11. Y. Li, P. C. Humphreys, G. J. Mendoza, and S. C. Benjamin, Resource Costs for Fault-Tolerant Linear Optical Quantum Computing, *Phys. Rev. X* **5**, 041007 (2015).

12. M. Pont, G. Corrielli, A. Fyrillas, I. Agresti, G. Carvacho, N. Maring, P.-E. Emeriau, F. Ceccarelli, R. Albiero, P. H. Dias Ferreira, N. Somaschi, J. Senellart, I. Sagnes, M. Morassi, A. Lemaître, P. Senellart, F. Sciarrino, M. Liscidini, N. Belabas, R. Osellame, High-fidelity four-photon GHZ states on chip. *npj Quantum Inf* **10**, 50 (2024).

13. R. Uppu, F. T. Pedersen, Y. Wang, C. T. Olesen, C. Papon, X. Zhou, L. Midolo, S. Scholz, A. D. Wieck, A. Ludwig, P. Lodahl, Scalable integrated single-photon source, *Sci. Adv*. **6**, eabc8268 (2020).

14. J. B. Sping, B. J. Metcalf, P. C. Humphreys, et al. "Boson sampling on a photonic chip", *Science* **339**, 798 (2013).

15. S. Bartolucci, P. Birchall, H. Bombín, *et al*. "Fusion-based quantum computation". *Nat Commun* **14**, 912 (2023).

16. J. Zhang, S. Chattaraj, S. Lu, and A. Madhukar, Mesa top quantum dot single photon emitter arrays: growth, optical characteristics, and the simulated optical response of integrated dielectric nanoantenna-waveguide system , *Jour. App. Phys.* **120** (2016).

17. J. Zhang, S. Chattaraj, Q. Huang, L. Jordao, S. Lu and A. Madhukar, On chip scalable highly pure and indistinguishable single photon sources in ordered arrays: Path to Quantum Optical Circuits. *Sci. Adv.* **8**, eabn9252 (2022).

18. J. Zhang, Q. Huang, L. Jordao, S. Chattaraj, S. Lu, A. Madhukar, Planarized spatially-regular arrays of spectrally uniform single quantum dots as on-chip single photon sources for quantum optical circuits. *APL Photonics* **5**, 116106 (2020).

19. Q. Huang, L. Jordao, S. Lu, S. Chattaraj, J. Zhang, and A. Madhukar, *Large-Area Spatially Ordered Mesa Top Single Quantum Dots: Suitable Single Photon Emitters for On-Chip Integrated Quantum Information Processing Platforms*, arXiv:2312.15132 (2023).

20. Q. Huang, L. Jordao, S. Lu, S. Chattaraj, J. Zhang, and A. Madhukar, *Spatially Ordered Spectrally Compliant On-Demand Scalable Quantum Emitter Large Arrays for Multi-Emitter Based Quantum Networks*, in *Quantum 2.0 Conference and Exhibition* (Optica Publishing Group, Rotterdam, 2024), p. QTh4A.1.

21. P. Tighineanu, R. Daveau, E. H. Lee, J. D. Song, S. Stobbe, P. Lodahl, Decay dynamics and exciton localization in large GaAs quantum dots grown by droplet epitaxy. *Phys. Rev. B* **88**, 155320 (2013).

22. S. Stobbe, T. W. Schlereth, S. Höfling, A. Forchel, J. M. Hvam, P. Lodahl, Large quantum dots with small oscillator strength. *Phys. Rev. B*, **82**(23), (2010).

23. A. Madhukar, K. C. Rajkumar, P. Chen, "In situ approach to realization of three-dimensionally confined structures via substrate encoded size reducing epitaxy on nonplanar patterned substrates", *Appl. Phys. Lett.* **62**, 1547 (1993).

24. Madhukar, A. Growth of semiconductor heterostructures on patterned substrates: defect reduction and nanostructures. *Thin Solid Films.* **231**, 8 (1993)

25. H. Larocque, M. A. Buyukkaya, C. Errando-Herranz, C. Papon, S. Harper, M. Tao, J. Carolan, C.-M. Lee, C. J. K. Richardson, G. L. Leake, D. J. Coleman, M. L. Fanto, E. Waks, D. Englund, Tunable quantum emitters on large-scale foundry silicon photonics. *Nat Commun* **15**, 5781 (2024).

26. T. Stolz, H. Hegels, M. Winter, B. Röhr, Y.-F. Hsiao, L. Husel, G. Rempe, and S. Dürr, "Quantum-logic gate between two optical photons with an average efficiency above 40%," *Phys. Rev. X* **12**, 021035 (2022).

27. E. Hanamura, "Very large optical nonlinearity of semiconductor microcrystallites," *Phys. Rev. B*, **37**(3), 1273–1279, (1988).

28. T. Takagahara, Electron-phonon interactions and excitonic dephasing in semiconductor nanocrystals. *Phys. Rev. Lett.* **71**, 3577 (1993).

29. H. Wang, Z.C Duan, Y.H. Li, S. Chen, J.P. Li, Y.M. He, M.C. Chen, Y. He, X. Ding, C.Z. Peng, C. Schneider, M. Kamp, S. Höfling, C.Y. Lu, J.W. Pan, Near-transform-limited single photons from an efficient solid-state quantum emitter. *Phys. Rev. Lett*, **116**, 213601 (2016).

30. A. Thoma, P. Schnauber, M. Gschrey, M. Seifried, J. Wolters, J.H. Schulze, A. Strittmatter, S. Rodt, A. Carmele, A. Knorr, T. Heindel, and S. Reitzenstein, Exploring dephasing of a solid-state quantum emitter via time-and temperature-dependent Hong-Ou-Mandel experiments. *Phys. Rev. Lett*, **116**, 033601 (2016).

31. A. Reigue, J. Iles-Smith, F. Lux, L. Monniello, M. Bernard, F. Margaillan, A. Lemaitre, A. Martin, D.P. McCutcheon, J. Mørk, R. Hostein, and V. Voliotis, Probing electron-phonon interaction through two-photon interference in resonantly driven semiconductor quantum dots. *Phys. Rev. Lett*, **118**, 233602 (2017).

32. S. Gerhardt, J. Iles-Smith, D.P. McCutcheon, Y.M. He, S. Unsleber, S. Betzold, N. Gregersen, J. Mørk, S. Höfling, C. Schneider, Intrinsic and environmental effects on the interference properties of a high-performance quantum dot single-photon source. *Phys. Rev. B* **97**, 195432 (2018)

33. X. Ding, Y. He, Z.C. Duan, N. Gregersen, M.C. Chen, S. Unsleber, S. Maier, C. Schneider, M. Kamp, S. Höfling, C.Y. Lu, J-W. Pan, On-demand single photons with high extraction efficiency and near-unity indistinguishability from a resonantly driven quantum dot in a micropillar. *Phys. Rev. Lett,* **116**(2), 020401, (2016).

34. J. Iles-Smith, D.P. McCutcheon, A. Nazir, and J. Mørk, Phonon scattering inhibits simultaneous near-unity efficiency and indistinguishability in semiconductor single-photon sources. *Nat. Photonics*, **11**, 521-526 (2017).

35. N.K. Langford, T.J. Weinhold, R. Prevedel, K.J. Resch, A. Gilchrist, J.L. O'Brien, G.J. Pryde and A.G. White. "Demonstration of a Simple Entangling Optical Gate and Its Use in Bell-State Analysis", *Phys. Rev. Lett.* **95**, 210504 (2005).

36. N. Kiesel, C. Schmid, U. Weber, R. Ursin and H. Weinfurter. "Linear Optics Controlled-Phase Gate Made Simple", *Phys. Rev. Lett.* **95**, 210505 (2005).

37. R. Okamoto, H.F. Hofmann, S. Takeuchi, and K. Sasaki. Demonstration of an Optical Quantum Controlled-NOT Gate without Path Interference, *Phys. Rev. Lett.* **95**, 210506 (2005)

38. H.F. Hofmann, Complementary Classical Fidelities as an Efficient Criterion for the Evaluation of Experimentally Realized Quantum Operations, *Phys. Rev. Lett.* **94,** 160504 (2005).

39. Altepeter, J., Jeffrey, E. and Kwiat, P. Photonic state tomography. *Adv. At. Mol. Opt. Phys*. **52**, 105–159 (2005).

40. C. H. Bennett, G. Brassard, S. Popescu, B. Schumacher, J. A. Smolin, and W. K. Wootters, Purification of Noisy Entanglement and Faithful Teleportation via Noisy Channels, *Phys. Rev. Lett.* **76**, 722 (1996).